\documentclass[9pt,twocolumn,twoside]{osajnl}

\journal{ao} 

\setboolean{shortarticle}{false}

\title{Polarimetric characterization of segmented mirrors}

\author[1]{A. Pastor Yabar}
\author[2,3]{A. Asensio Ramos}
\author[4]{R. Manso Sainz}
\author[2,3]{M. Collados}

\affil[1]{Institute for Solar Physics, Dept. of Astronomy, Stockholm University, AlbaNova University Centre, 106 91 Stockholm, Sweden}
\affil[2]{Instituto de Astrof\'{\i}sica de Canarias (IAC), c/ V\'{\i}a L\'actea s/n, La Laguna, Tenerife, Spain, 38205}
\affil[3]{Departamento de Astrof\'{\i}sica, Universidad de La Laguna, Avda, Astrof\'{\i}sico S\'anchez, s/n, La Laguna, Tenerife, Spain, 38206}
\affil[4]{G\"ottingen, Germany}

\affil[*]{Corresponding author: adur.pastor@astro.su.se}

\begin{abstract}
We study the impact of the loss of axial symmetry around the optical axis on the
polarimetric properties of a telescope with segmented primary mirror when each segment 
is present in a different aging stage. The different oxidation stage of each segment
as they are substituted in time leads to non-negligible cross-talk
terms. This effect is wavelength dependent and it is mainly determined by the properties
of the reflecting material. For an aluminum coating, the worst polarimetric behavior due to
oxidation is found for the blue part of the visible. Contrarily, dust ---as modeled in this work--- does not significantly
change the polarimetric behavior of the optical system . Depending 
on the telescope, there might be segment substitution sequences that strongly attenuate this instrumental polarization.
\end{abstract}

\setboolean{displaycopyright}{false}

\begin{document}

\maketitle

\section{Introduction}
The new generation of large telescopes, such as the European Extremely Large Telescope
(\mbox{E-ELT}) \cite{gilmozzi2007} or the Thirty Meter
 Telescope (TMT)\cite{sanders2013} plan gigantic and segmented primary surfaces.
 As an example, the collecting area of the \mbox{E-ELT} primary mirror is larger than a basketball court.
 Maintaining the homogeneity of such a huge optical surface is a big challenge ---perhaps
an impossible one---. Removing, aluminizing and installing each of its 798 hexagonal
segments may take over a year and by then, there will be a significant degradation
on the first ones. One consequence of an inhomogeneous primary mirror is simply
a reduced reflectivity. Another potentially more critical effect might be its
impact on polarization and the polarimetric properties of the telescope.
Polarization is of special relevance from a diagnostic
 point of view because it is intimately related with any symmetry breaking
 phenomenon taking place on the object of interest or the presence of magnetic fields.
 Thus it is possible to study, among others, the magnetic dynamo in the Sun and other cool stars
 \cite{brun2017}, the spatial geometry of (either magnetic or non-magnetic)
 physical systems such as planetary nebulae \cite{martinezgonzalez2015}, even
 spatially unresolved stars opening up the possibility of carrying out stellar imaging
 \cite{kochukhof2016}, the study of active galactic nuclei \cite{carnerero2017}
 or the presence and properties of extra-solar planets \cite{berdyugina2008},
 with specific geometrical configurations that give rise to specific
 polarization states. Finally, there is increasing bibliography on the potential
 applicability of the polarization of the light in order to find extraterrestrial
 life \cite{kiang2018}.

Oblique reflection in an optical surface polarizes (linearly, perpendicularly to the
plane of reflection) the incoming light \cite{capitani1989}. Most
telescopes have primary and secondary mirrors in an axially symmetric configuration.
Their net instrumental polarization (along the optical axis) is zero 
because two points in the mirror located at the same distance from
the optical axis but $90^{\circ}$ apart should polarize to the same degree but with
opposite sign. In fact, this property can be used on the whole optical path
to design telescopes that are free of instrumental polarization, such as the European 
Solar Telescope (EST)\cite{collados2010}.
Large 10-meter class telescopes with segmented hexagonal mirrors are not axially symmetric
but the 6-fold symmetry of their mirrors should also lead to perfect cancellation after 
integration over the whole mirror.

Also, in smaller telescopes, the angles of incidence are very close to normal which
severely limits the amount of instrumental polarization generated by the primary
and secondary mirrors. Reflections on the external parts of the primary mirror
of the \mbox{E-ELT} can reach angles $\sim 15^{\circ}$ 
which may yield non-negligible polarization effects.

In addition, in telescopes with segmented primary mirrors, the maintenance 
of these elements is usually done in a sequential manner so that the scientific
availability of the facility is maximized. Thus this process can introduce an
additional source of symmetry loss in the optical system. Polarimetry in telescopes
with segmented mirrors is already feasible (for instance CanariCam \cite{telesco2003})
and will be more frequent soon as the incoming 
new generation instrumentation for the Gran Telescopio Canarias (GTC) ---MIRADAS \cite{eikenberry2016}--- 
as well as part of the instrumentation to be installed in the future \mbox{E-ELT}
are planning to acquire high sensitivity polarimetric data. It is thus of interest
to evaluate the potential instrumental polarization that symmetry loss due to 
sequential segment substitution might introduce.

In this note we study the severity of the intrinsic inhomogeneity of the primary
mirror in large segmented telescopes in general and in two cases of practical
interest in particular ---\mbox{E-ELT} and GTC---, with
very different segment-to-mirror area ratios. In contrast to previous works
\cite{jennings2002, anche2018, anche2019}, here we consider variations on
the reflectivity of the segments due to oxidation of the aluminum (Al)
and dust deposition characterizing the instrumental polarization of the
primary and secondary mirrors. This study is valuable to understand the
contribution to the global budget of instrumental polarization of the telescope,
to design polarimetric calibration strategies to compensate for it, and to propose
the most convenient segment replacement method to mitigate adverse effects.

\section{Method}
\label{sect:methods}

In this work we have developed a ray-tracing numerical code that evaluates the
polarimetric properties of the segmented primary and secondary mirrors of two
different telescopes: GTC and \mbox{E-ELT}, as we assume that we
are dealing with plane waves (for more complex processes a more suitable
formalism \cite{yun2011} could be employed). In particular, we investigate the effect
on the polarimetric properties of the optical system as the axial symmetry
with respect to the optical axis is broken either by the shape of the mirror
itself or additional sources that might contribute to the loss of axial
symmetry. To do so, we use the Stokes formalism, which allows to fully 
characterize the polarimetric properties of an optical system. 

\subsection{Reflecting surfaces}
\label{subsect:mueller_reflection}

\begin{figure}
    \centering
    \includegraphics[width=0.5\textwidth]{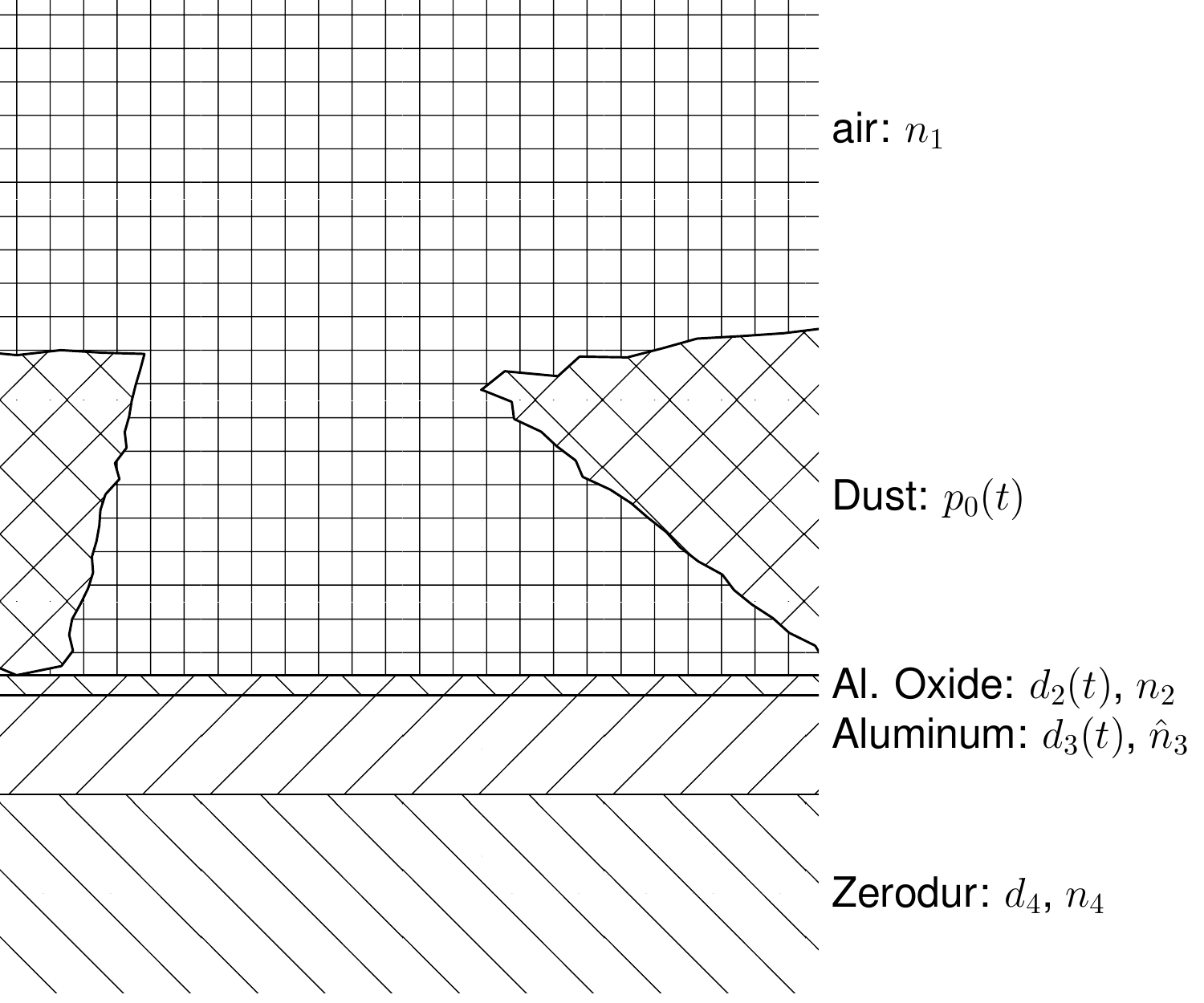}
    \caption{Elements of the primary reflecting surface: dust, aluminum oxide, aluminum,
    and Zerodur for a given segment at a timestep $t$.}
    \label{fig:reflectingsurfaceprimary}
\end{figure}

In the Stokes formalism, the action of optical devices is described by linear
transformations over the Stokes pseudo-vector, i.e. by means of $4\times4$
matrices. The reflection on a mirror in this formalism is given by
the following transformation matrix \cite{born1999}:

\begin{equation}
\mathfrak{M}=\frac{1}{2}
\left(\begin{array}{cccc}
t_{1}&t_{2}&0&0\\
t_{2}&t_{1}&0&0\\
0&0&t_{3}\cos\delta&t_{3}\sin\delta\\
0&0&-t_{3}\sin\delta&t_{3}\cos\delta
\end{array}\right),
\label{Eq:1 muellerreflectionmatrix}
\end{equation}

\noindent where $t_{1}=|\rho^{\parallel}|^{2}+|\rho^{\perp}|^{2}$,
$t_{2}=|\rho^{\parallel}|^{2}-|\rho^{\perp}|^{2}$,
$t_{3}=2\rho^{\parallel}\rho^{\perp}$, and
$\delta=\phi^{\perp}-\phi^{\parallel}$. The symbols $\rho$ and $\phi$ refer to the
modulus and phase, respectively, of the Fresnel coefficients for reflection $r^{\parallel}$ and $r^{\perp}$:

\begin{gather}
    r^{\parallel}=
    \rho^{\parallel}e^{i\phi^{\parallel}}
    \\
    r^{\perp}=
    \rho^{\perp}e^{i\phi^{\perp}}
    .
\end{gather}

The expressions of the Fresnel coefficients depend upon the specific composition of the
surface in which light is reflected. For the primary mirror, the reflecting surface consists of the
substrate of the mirror, the conductor, which is deposited above it, the oxide
that forms above the conductor due to the action of the atmosphere and the dust that
accumulates on top (Fig. \ref{fig:reflectingsurfaceprimary}). Except for dust,
which we account for in a different way, the first three layers can be safely modeled
by means of the thin film theory. According to this theory, the Fresnel
coefficients do only depend on the thickness and the refractive indices (real
or complex) of the films that form the reflective surface. We use the following formula
\cite{born1999} for their computation:

\begin{equation}
r^{\parallel}=
\frac{(m_{11}+m_{12}\,p_{4})p_{1}
-(m_{21}+m_{22}\,p_{4})}
{(m_{11}+m_{12}\,p_{4})p_{1}
+(m_{21}+m_{22}\,p_{4})},
\label{Eq:fresnelcoefficentthinfilms}
\end{equation}
where $p_{\rm k}=n_{\rm k}\cos\theta_{\rm k}$, with $k=1$ for air, $k=2$ for the oxide film, $k=3$ for
the conductor and $k=4$ for the substrate. A similar expression is obtained for the
perpendicular Fresnel coefficient 
($r^{\perp}$) by using $q_{\rm k}=\frac{\cos\theta_{\rm k}}{n_{\rm k}}$ instead of
$p_{\rm k}$ in Eq. (\ref{Eq:fresnelcoefficentthinfilms}).
Eq. (\ref{Eq:fresnelcoefficentthinfilms}) has explicit dependence
on both the air and substrate, while the intermediate layers enter through the characteristic matrix of
the reflecting surface. 
The coefficients $m_{lm}$ of this matrix are given by:
\begin{equation}
M=\prod_{k=4}^{1}M_{\rm k}=\left(\begin{array}{cc}m_{11}
&m_{12}\\m_{21}&m_{22}\end{array}\right).
\end{equation}
Each one of the characteristic matrices is given by the expression:
\begin{equation}
M_{\rm k}=
\left(\begin{array}{cc}
\cos\beta_{\rm k}&-\frac{i}{p_{\rm k}}\sin\beta_{\rm k}\\
-i\,p_{\rm k}\sin\beta_{\rm k}&\cos\beta_{\rm k}
\end{array}\right),
\label{eq:characteristic_matrix}
\end{equation}

where $\beta_{\rm k}=k_{0}\,n_{\rm k}\,d_{\rm k}\,\cos\theta_{\rm k}$, with $k_{0}$ being the 
wavenumber of the radiation, $n_{\rm k}$ being the refraction index of the $k$-th layer (either
real or complex), $d_{\rm k}$ the thickness of the $k$-th layer and $\theta_{\rm k}$ 
the angle of incidence on the $k$-th layer.

The dust layer cannot be treated within the thin film formalism because the sizes
of typical dust particles are already of the order of the wavelength of the light
under consideration (typically in the visible or near infrared). The model that
we choose is discussed in Sect. \ref{subsect:method_dust}.

In order to simplify the interpretation of the results, we consider a secondary mirror made of
a substrate layer over which we lay a conductor. No dust is considered since its
deposition is less probable than on the primary mirror. Additionally, we keep the thickness of the
conductor constant. Therefore, any variation of the polarization properties of the
telescope is then associated with changes in the primary mirror. 

With the previous theory, once the wavelength of the incoming radiation, the incidence angle and the
thickness and reflecting indices of the various films are given, the calculation of the
Fresnel coefficients for any reflection in the optical system is possible. This allows us to calculate 
the reflective properties for any incident ray of the system.

\subsection{Geometrical setup}

\begin{table}
\caption{Radius of curvature $R$ and conic constant $K$ for primary ($^{\rm prim}$) and
secondary ($^{\rm sec}$) mirror of GTC and \mbox{E-ELT} considered. $d$ is the distance between
the primary and secondary mirrors at the optical axis.}
\label{Tab1:conicparameters}
\centering                                      
\begin{tabular}{l c c c}
\hline\hline
Parameter & GTC & \mbox{E-ELT} & Units \\
\hline
 $R_{c}^{\rm prim}$  & 33.0 & 69.168 & m \\
 $K^{\rm prim}$      & -1.002250 & -0.993295 & - \\
 $R_{c}^{\rm sec}$   & 3.899678 & 9.313 & m \\
 $K^{\rm sec}$       & -1.504835 & -2.28962 & - \\
 $d$       & 14.73941 & 31.415 & m \\
\hline                                             
\end{tabular}
\end{table}

\begin{figure*}
    \centering
    \includegraphics[width=0.55\textwidth]{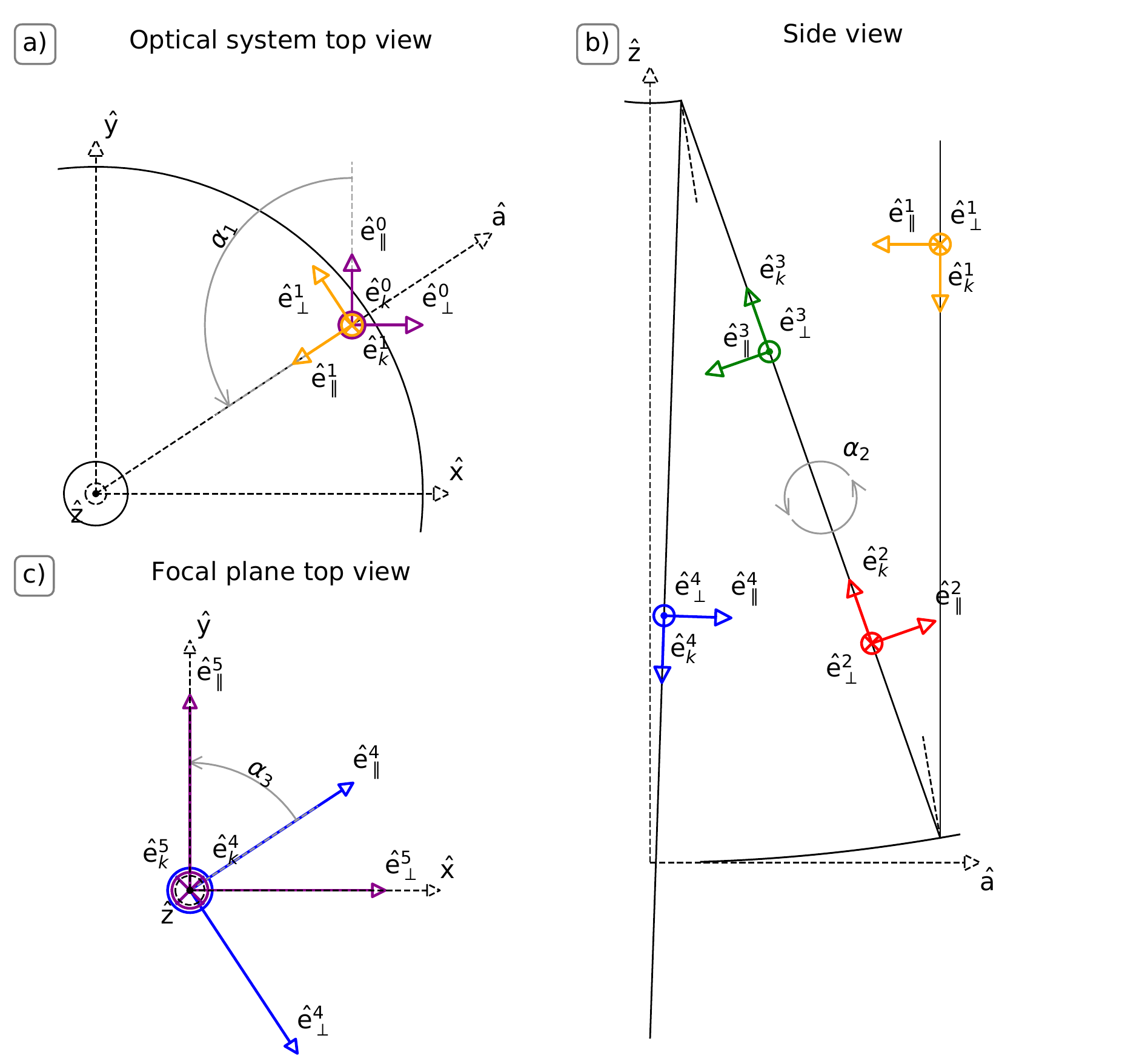}
    \caption{Panel a) shows the polarization reference system ($\hat{\rm e}^{0}$, in purple
    color) that has to be rotated by $\alpha_{1}$, which depends upon the $x$ and $y$ position
    of the ray, in order to get the reference system ($\hat{\rm e}^{1}$ in orange) so that
    reflection can be written as in Eq. (\ref{Eq:1 muellerreflectionmatrix}). After that first
    rotation, there are two reflections and one additional rotation ($\alpha_{2}$) that take
    in the same fashion once in the $\hat{\rm a}-\hat{\rm z}$ plane, as depicted in panel b).
    Finally, panel c) shows the last rotation ($\alpha_{3}$) required in order to set the
    polarimetric properties of each point of the optical system in a common reference
    system ($\hat{\rm e}^{5}\equiv\hat{\rm e}^{0}$).}
    \label{fig:telescope_optical_layout}
\end{figure*}

\begin{figure}
    \centering
    \includegraphics[width=0.44\textwidth]{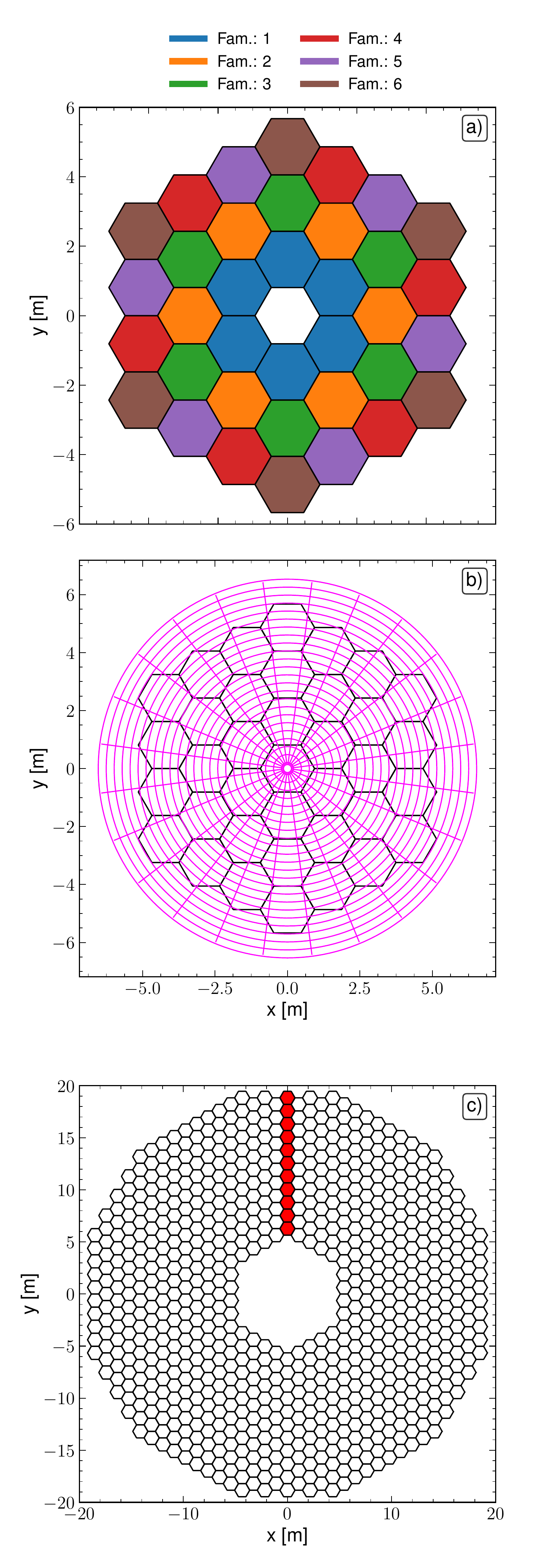}
    \caption{a) GTC primary mirror layout color coded according to the family of 
    segments they belong to. b) Ray quadrature layout over GTC's primary mirror
    layout. c) \mbox{E-ELT}'s primary mirror layout with the segments used in Fig.
    \ref{fig:muellersegmentdistance} highlighted in red.}
    \label{fig:primary_mirrors_quadrature}
\end{figure}

In order to know the angle of incidence needed for the calculation
of $\beta_{\rm k}$ in Eq. (\ref{eq:characteristic_matrix}) for both the primary and secondary mirrors, 
we need to know the inclination of the ray of light entering the telescope with respect to the
optical axis. This parameter is an input supplied by the user, and in this work
it will always be aligned with the optical axis of the system. For the numerical
calculation, we consider many incoming light beams that reach
the primary mirror in various positions. Panel a) of Fig. 
\ref{fig:telescope_optical_layout} shows (from a top view of the telescope) one potential
case of such an incident ray. The reference system for the optical components is chosen
so that $z$ is along the optical axis of the system and $x$ and
$y$ in the plane perpendicular to it. $z=0$ is in the
intersection of the optical axis with the surface described by the primary mirror
(see for instance the side view of the system in panel b) of Fig. \ref{fig:telescope_optical_layout}).
To fully take into account the symmetry of
the problem, the $(x,y)$ coordinates for the rays
are defined following a quadrature in polar coordinates so that
we ensure that every ray has its complementary one (i.e. at exactly
the same distance but at 90$^{\circ}$) if it is possible. An example  
of a very coarse quadrature for GTC is shown in panel b) of 
Fig. \ref{fig:primary_mirrors_quadrature}. In the figure,
for each of the area elements (highlighted in magenta), we assume that the
polarimetric properties are the ones given for the central point of that small
area. Note here that, for the sake of clarity, this example shows the case for 240
rays, a very small number of rays. We also show all quadrature points, independently
of whether they reach the focal plane of the telescope. Some of these rays will not
reflect on the primary or on the secondary mirror, so they are excluded in the calculation.

In addition to the incident position of each ray, the calculation of the angle
of incidence requires the knowledge of the specific shape of the primary mirror. For
the telescopes here considered, the primary mirror is formed by gathering hexagonal 
segments (as seen from the
optical axis). In Fig. \ref{fig:primary_mirrors_quadrature} we show the segment
distribution for each telescope: in panel a) for GTC in colored format and
in panel c) for \mbox{E-ELT}. These are built following the specifications in the
conceptual design for GTC (defined on 1997) and the construction proposal for the \mbox{E-ELT}
(defined on 2011). They contain 36 and 798 hexagonal segments for the primary mirror, respectively.
The segment side side size is 936 mm and 725 mm, respectively. A certain amount of hexagons are
not present in the center so that each mirror keeps an approximate internal circular area
with a radius of 0.4 and 4.7 m for GTC and \mbox{E-ELT}, respectively. The \mbox{E-ELT}
secondary mirror is of circular shape and has a radius of 4.2 m. The secondary
mirror of GTC is serrated with a maximum radius of 1.1766 m and an internal free
area of hexagonal shape with a side size of 0.1388 m. In addition to the shape
of the mirrors, we also need their curvature which, for the primary, is given by:

\begin{equation}
x^2+y^2-2\,R\,z+(1+K)\,z^{2}=0,
\end{equation}
where the values for $R$ and $K$ are in Table \ref{Tab1:conicparameters} for
both telescopes. Combining all these parameters one can determine the angle of incidence of each
ray over the primary mirror. Once this reflection takes place, the light is
reflected to the secondary mirror whose shape is given by:
\begin{equation}
x^2+y^2-2\,R\,\zeta+(1+K)\,\zeta^{2}=0\label{Eq:conic},\\
\end{equation}
with $\zeta=z-d$, being $d$ the distance between both mirrors along the optical
axis. The actual values for the conic constant and the radius of curvature of
the secondary mirror are also shown in Table \ref{Tab1:conicparameters}. 

With this information we can compute the angle of incidence for each ray both
in the primary and secondary mirrors. Thus, in order to calculate the reflection
matrices described in Sect. \ref{subsect:mueller_reflection} we need the
thickness and refractive indexes of each layer considered.

\subsection{Thickness and refractive index}
\label{subsect:method_indexrefraction}

The aging of the various primary mirror segments requires taking
into account the time variation of the conductor
and oxide films thickness. As mentioned before the dust layer is handled differently and discussed in 
Sect. \ref{subsect:method_dust}. Concerning the oxide thickness,
we assume an exponential growth law. This is expected since at the beginning the
conductor is directly exposed to air and so the oxidation is fast. Once the
oxide film gets thicker, this oxide layer itself prevents the conductor from the
contact with air and, consequently, the growth rate of the oxide film is strongly reduced.
This way, the thickness of the oxide layer with time is assumed to be given by:

\begin{equation}
    d_{2}=d_\mathrm{ox}^\mathrm{max}\,(1-e^{-t/\tau_{\rm ox}}),
\end{equation}
where $d_\mathrm{ox}^\mathrm{max}$ is the maximum thickness allowed for the oxide layer, set
here to 0.1 $\mu$m, $\tau_{\rm ox}$ is the growth rate for the oxide film,
that we choose to be 1349 days, and $t$ is the time, measured in days, since the
segment exchange. This behavior is complementary to that of the conductor (namely
$d_{3}=d_{\mathrm{al}}^{0}-d_{2}$), for which we require an initial thickness:
$d_{\mathrm{al}}^{0}=1$ $\mu$m.

\begin{figure}
    \centering
    \includegraphics[width=0.44\textwidth]{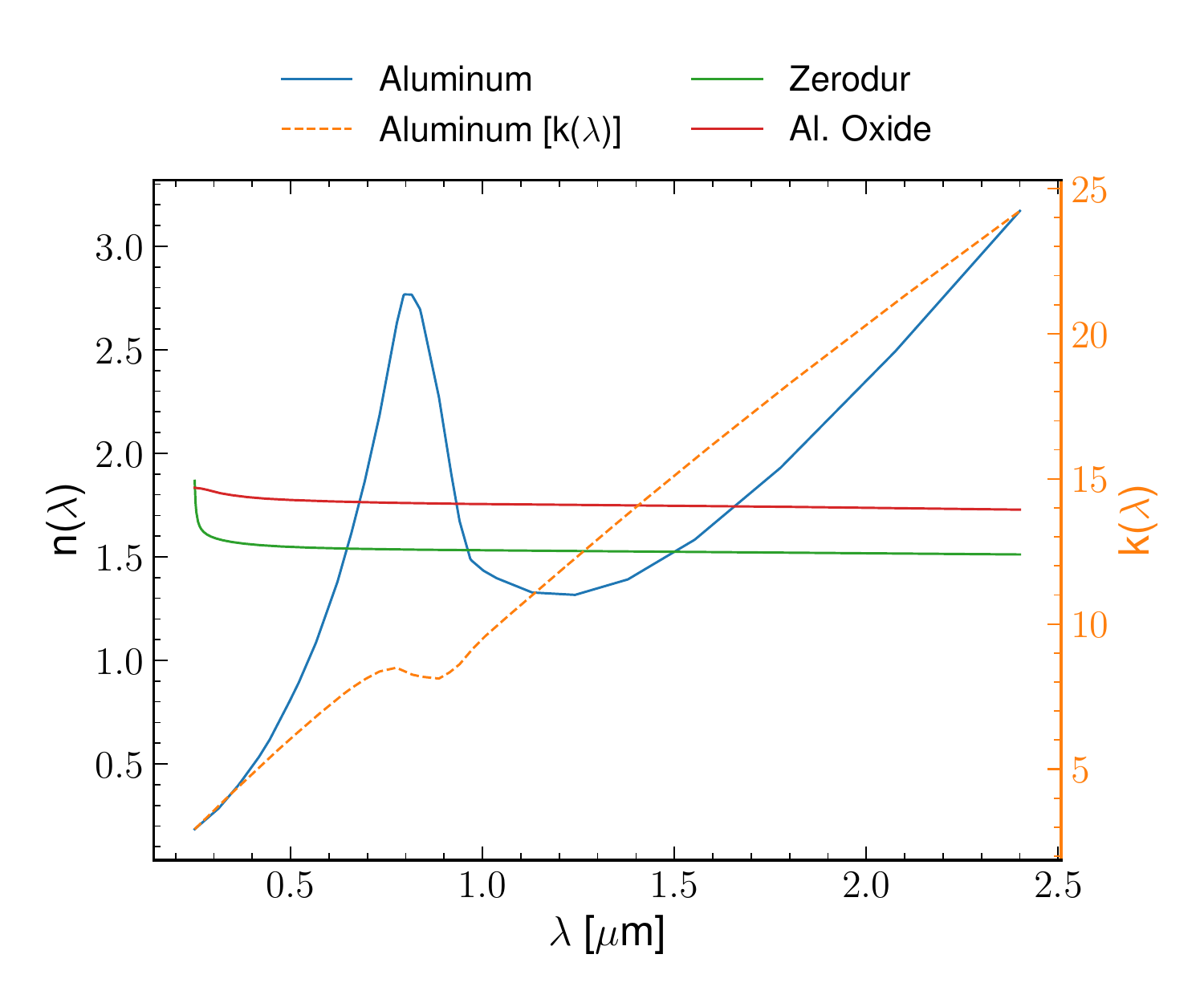}
    \caption{Refractive index's real part (solid line) and imaginary part (dashed
    line) for the materials employed and their wavelength dependence.}
    \label{fig:reflection_indexes}
\end{figure}

In addition to the thickness of each layer, the calculation of the Mueller
matrix requires the refractive index of each material, together with
the wavelength variation (shown in Fig. \ref{fig:reflection_indexes}). In this
work, we assume that all mirrors are covered with aluminum (whose refractive index
is complex), the oxide film is then aluminum oxide
($\text{Al}_{2}\text{O}_{3}$), the substrate is Zerodur.

\subsection{Mueller matrix for each segment}

With this modeling, it is possible to calculate the reflection matrix for
each ray in both the primary and secondary mirrors. To do so, we have to take into account that the
Mueller matrix for reflection as shown in Eq. (\ref{Eq:1
muellerreflectionmatrix}) is only valid when the vector 
$\hat{e}_{\parallel}$ lies in the incidence-reflection plane.
Since this is not the
case in general, one has to include some additional rotations, which vary from ray to
ray. These rotations ensure that the necessary conditions for validity 
of the reflection matrix are satisfied. The train of rotation is
as follows: i) a first rotation ($\alpha_{1}$) transforms the system
from the global reference frame for 
the Stokes parameters ($\hat{\rm e}^{0}$) to the one preferred for computing the
reflection in each point of the primary mirror ($\hat{\rm e}^{1}$ see panel a) in
Fig. \ref{fig:telescope_optical_layout}), ii) a second rotation ($\alpha_{2}$) takes
the outgoing reference frame from the first reflection ($\hat{\rm e}^{2}$)
to the preferred one for the second reflection ($\hat{\rm e}^{3}$, panel b) in the figure), and iii) a
third rotation ($\alpha_{3}$) takes the reference frame from the second reflection 
($\hat{\rm e}^{4}$) back to the global
reference frame ($\hat{\rm e}^{5}$ in panel c) of the figure). Thus, for each ray, the equivalent Mueller matrix of 
the two-mirror model here considered is given by: 

\begin{equation}
\mathfrak{M}_{x,y}=R_{x,y}(\alpha_{3})\,\mathfrak{M}_{\rm x,y}^{\rm sec}\, 
R_{x,y}(\alpha_{2})\,\mathfrak{M}_{x,y}^{\rm prim}\, R_{\rm x,y}(\alpha_{1}),
\end{equation}
where $R_{\rm x,y}(\alpha_{1}=\pi/2+\arctan(y/x)
)$ is the
first rotation matrix for the
point $x,y$ on the primary mirror (see panel a in Fig.
\ref{fig:telescope_optical_layout}), $\mathfrak{M}_{\rm x,y}^{\rm prim}$ is the
reflection matrix for that point in the primary mirror, $R_{x,y}(\alpha_{2}=\pi)$ is the second
rotation that takes the reference frame from that for the primary to the one
needed for the secondary reflection, $\mathfrak{M}_{\rm x,y}^{\rm sec}$ takes
into account the
reflection on the secondary mirror, and $R_{\rm x,y}(\alpha_{3}=
\pi/2-\arctan(y/x)=\pi-\alpha_{1}
)$ finally rotates to the common reference frame for polarization.

\subsection{Dust}
\label{subsect:method_dust}

The last element to take into account is dust deposition on the mirrors
as well as its modeling.
We cannot use the thin-film theory because the typical size of a dust particle
is already similar or larger
than the wavelength for visible or near infrared light. In this work we model the
effect of dust from a statistical point of view rather than individually for
each ray beam. We assume that each segment is partly covered by dust. 
If a ray falls into a dusty area, then, by hypothesis that ray 
becomes part of stray light and disappears from the calculation.
This is represented in Fig. \ref{fig:reflectingsurfaceprimary} by a time-varying
dust particle coverage. Mathematically,
we propose that the incident light of each segment is modified by:

\begin{equation}
  \mathfrak{M}_{\rm dust}^{i}=(1-p^{i}_{0}(t))1\!\!1,
\end{equation}

where $1\!\!1$ is the identity matrix and $p^{i}_{0}$ is the probability of the
incoming light hitting dust particles and thus disappearing. This parameter
is segment-dependent in the sense that for a given time, each segment will
be characterized by a different value of $p_{0}$, which is function of the
time from the last segment exchange.

This way the effect of dust over each segment is described as a linear
transformation whose action precedes to any other optical transformation
, i.e. effectively it attenuates the incoming light reaching to each
segment:

\begin{equation}
\mathfrak{M}^\mathrm{final}_{x,y}=\mathfrak{M}_{x,y} \mathfrak{M}_{\rm dust}^i
\end{equation}

This is a very simplified approach for the effect of dust over an optical system
but it includes the potential effect over polarization due to symmetry breaking effects
produced by different amounts of dust in each segment. A more realistic model for
dust deposition and its effect over the overall transmission of the system is
out of the scope of this work.

\subsection{Replacement sequences}
\label{subsect:replacement_sequences}

\begin{figure}
    \centering
    \includegraphics[width=0.44\textwidth]{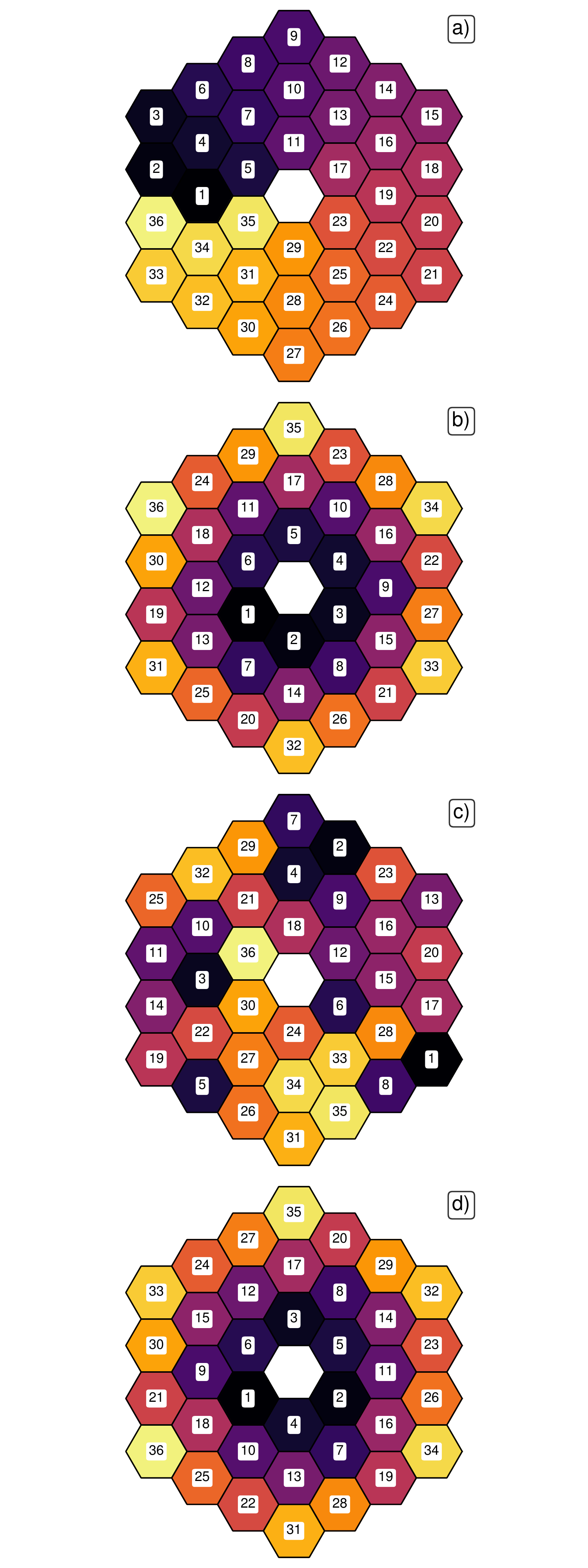}
    \caption{Segment substitution sequences for GTC: ``azimuth'' (panel a), ``linear'' (panel b), ``random'' (panel c), and ``symmetry'' (panel d). The label in each segment represents the position in the segment substitution sequence.}
    \label{fig:segment_substitution_sequences}
\end{figure}

With the model described above we can fully characterize the polarimetric properties
of any telescope (GTC and \mbox{E-ELT} in our case) for any given time. In order to study the effect that
differential aging of each segment has over the polarimetric properties of the
optical system, we consider different segment substitution strategies. Each
substitution sequence will take place along a whole year (approximately), mimicking the
substitution cadence that is already at work at GTC and the one expected for \mbox{E-ELT}. In the
first substitution method, which is labeled as ``azimuth'' (see index labelling in
panel a in Fig. \ref{fig:segment_substitution_sequences} for the specific order for
the case of GTC), we exchange the
segments according to their azimuth value (and for segments with the same azimuth
position, on the basis of their distance to the optical axis from the furthest
 to the closest). In the second case, labeled as
``linear'' (see panel b in Fig. \ref{fig:segment_substitution_sequences}), we exchange the segments according to their distance to the optical
axis and, for these at same distance to the optical axis, according to their
azimuth value. For the third case, labeled as ``random'' (see panel c in Fig. \ref{fig:segment_substitution_sequences}), we follow a random
order exchanging the segments. The fourth case is labeled ``symmetry'' (see panel d in Fig. \ref{fig:segment_substitution_sequences}), where the
segments are exchanged according to their distance to the optical axis and, for
these at same distance to the optical axis, the segments are exchanged in such a
way that two consecutive exchanges are done with a difference in the azimuth
of the segments of 120$^{\circ}$ or more.

Note that some of the sequences considered here are not actually feasible in practice.
It is so because the designs for GTC and \mbox{E-ELT} mentioned above require
a number of segments that share exactly the same shape, i.e. they are
interchangeable. Segments that share the same shape are referred here as
belonging to the same ``family''. Consequently, in
order to replace a segment and not leave a hole in the primary mirror it is
mandatory to have an additional segment for each of these families. The number
of families is 6 for GTC (see coloring in panel a of Fig. 
\ref{fig:primary_mirrors_quadrature}) and 133 for \mbox{E-ELT}. The planned cleaning strategy for \mbox{E-ELT}
requires that two segments are changed daily, so that the sequences labeled 
``linear'' and ``symmetry'' cannot be executed in practice, as it would require 
to have at least two additional segments per family instead of the only one 
planned. For GTC, any of the considered
cases is applicable, as the cleaning process of each segment takes less than two
days and segments are assumed to be changed every ten days. Also, the ``random''
sequence requires additional constraints because two mirrors of the same family
cannot be exchanged in consecutive timesteps. Despite this limitation for \mbox{E-ELT},
we keep the analysis for all the substitution sequences considered for both
telescopes to better visualize the importance of taking into account the
differential aging of segments when doing polarimetry with segmented primary
mirrors.

\section{Results and discussion}
We apply the polarimetric model with progressive complexity to the primary and
secondary mirror trying to identify the contribution the different model parameters have
on the polarimetric properties of the optical system.

\subsection{Ideal reflection: individual segments}

\begin{figure*}
    \centering
    \includegraphics[width=0.85\textwidth]{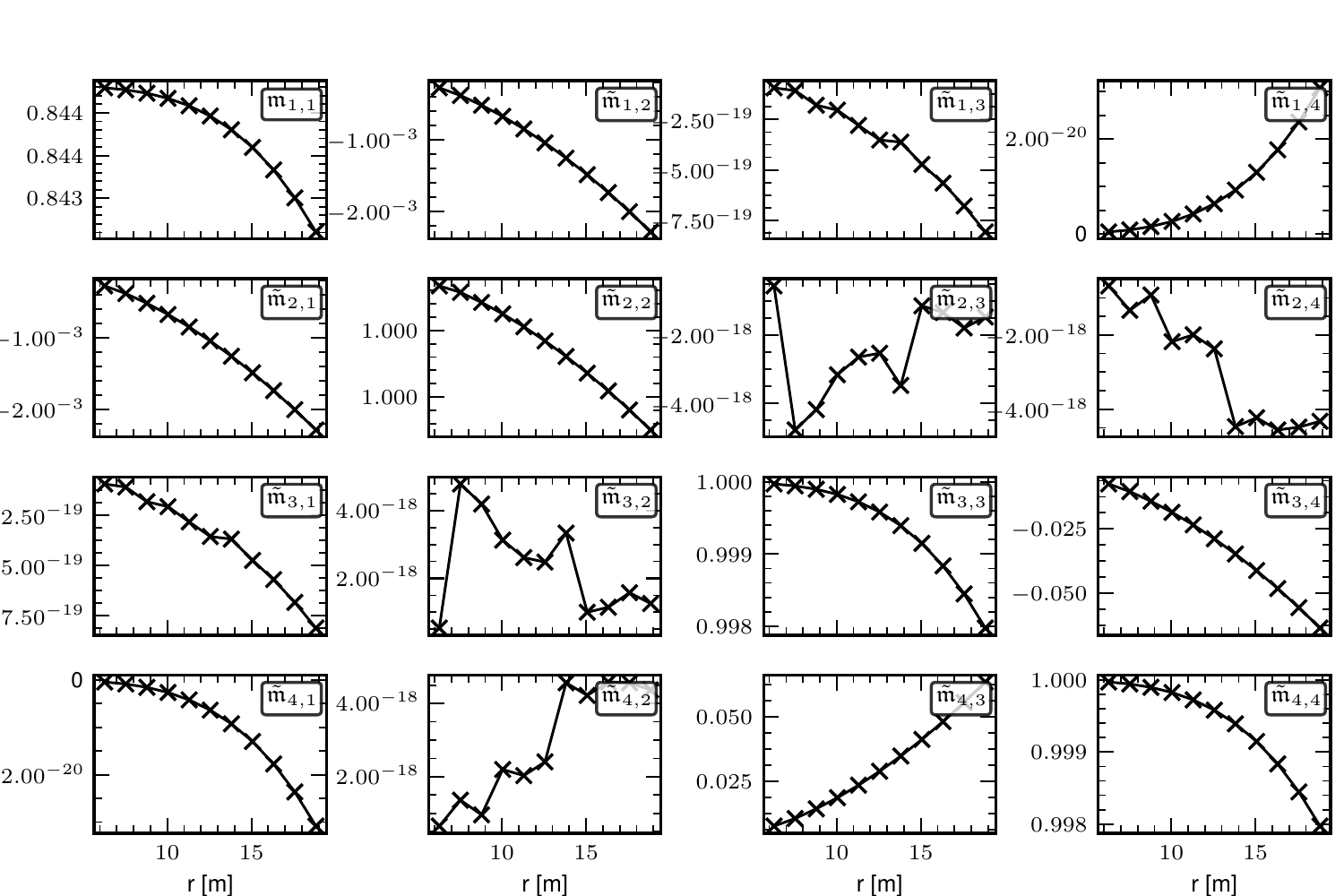}
    \caption{Mueller matrix for the eleven \mbox{E-ELT} segments highlighted in Fig.
    \ref{fig:primary_mirrors_quadrature} as a function of their distance to the optical
    axis at 500 nm. Note that elements other than $\mathfrak{m}_{1,1}$ represent
    the reflection corrected value, i.e. 
    $\tilde{\mathfrak{m}}_{i,j}=\mathfrak{m}_{i,j}/\mathfrak{m}_{1,1}$, for each distance value.}
    \label{fig:muellersegmentdistance}
\end{figure*}

Before proceeding to the polarimetric characterization of the optical system
as a whole, i.e. the integration over the whole surface area, it is worth
considering the polarimetric properties of individual segments as they allow
a more detailed insight into the system behavior. To do so, we consider the
various \mbox{E-ELT} segments highlighted in red in Fig. \ref{fig:primary_mirrors_quadrature}.
Figure \ref{fig:muellersegmentdistance} shows each Mueller matrix element value
for these segments as a function of the distance to the optical axis. That is
the only difference between the segments as here we are limiting ourselves to
an ideal reflector, i.e., neither dust nor oxidation is considered. Due to the
specific distribution of the segments (along the vertical direction) only
elements $\tilde{\mathfrak{m}}_{2,1}$, $\tilde{\mathfrak{m}}_{1,2}$, $\tilde{\mathfrak{m}}_{4,3}$
, and $\tilde{\mathfrak{m}}_{3,4}$ take non-negligible values (apart from diagonal elements)
as the segments are aligned with one of the polarization reference system. Note
that, in Fig. \ref{fig:muellersegmentdistance}, elements other than $\mathfrak{m}_{1,1}$ represent the reflection
corrected value, i.e. $\tilde{\mathfrak{m}}_{i,j}=\mathfrak{m}_{i,j}/\mathfrak{m}_{1,1}$.
The absolute amplitude of these Mueller elements increase as the segment distance
to the optical axis increases. Thus, the same reflecting surface of a segment
at two different distances from the optical will have a different impact on the
polarimetric behavior of the whole telescope.

\subsection{Ideal reflection: whole telescope}

Now we consider the polarimetric properties of the whole optical system in 
the same ideal configuration, i.e. without dust or oxidation.
In order to check this, we first look at the Mueller 
matrix for an ideal GTC using approximately $1.7\times10^{7}$
rays at 500 nm:

\begin{equation*}
\mathfrak{M}_{\rm GTC}^{\rm ideal}=\left(
\begin{array}{cccc}
0.844026 & 1^{-18} & 1^{-19} & 1^{-35} \\
1^{-18} & 0.844020 & 1^{-17} & 1^{-18} \\
1^{-19} & 1^{-17} & 0.844020 & 1^{-17} \\
1^{-34} & 1^{-18} & 1^{-17} & 0.844014
\end{array}
\right),
\end{equation*}

where the exponent $N$ is used to compactly express $\times10^{N}$ and the exact mantissa
for matrix elements with values smaller than $10^{-16}$ are omitted for clarity, as
they are essentially compatible with zero. And for the case of EELT, 
with approximately $2.8\times10^{8}$ rays at 500nm:

\begin{figure*}
    \centering
    \includegraphics[width=0.85\textwidth]{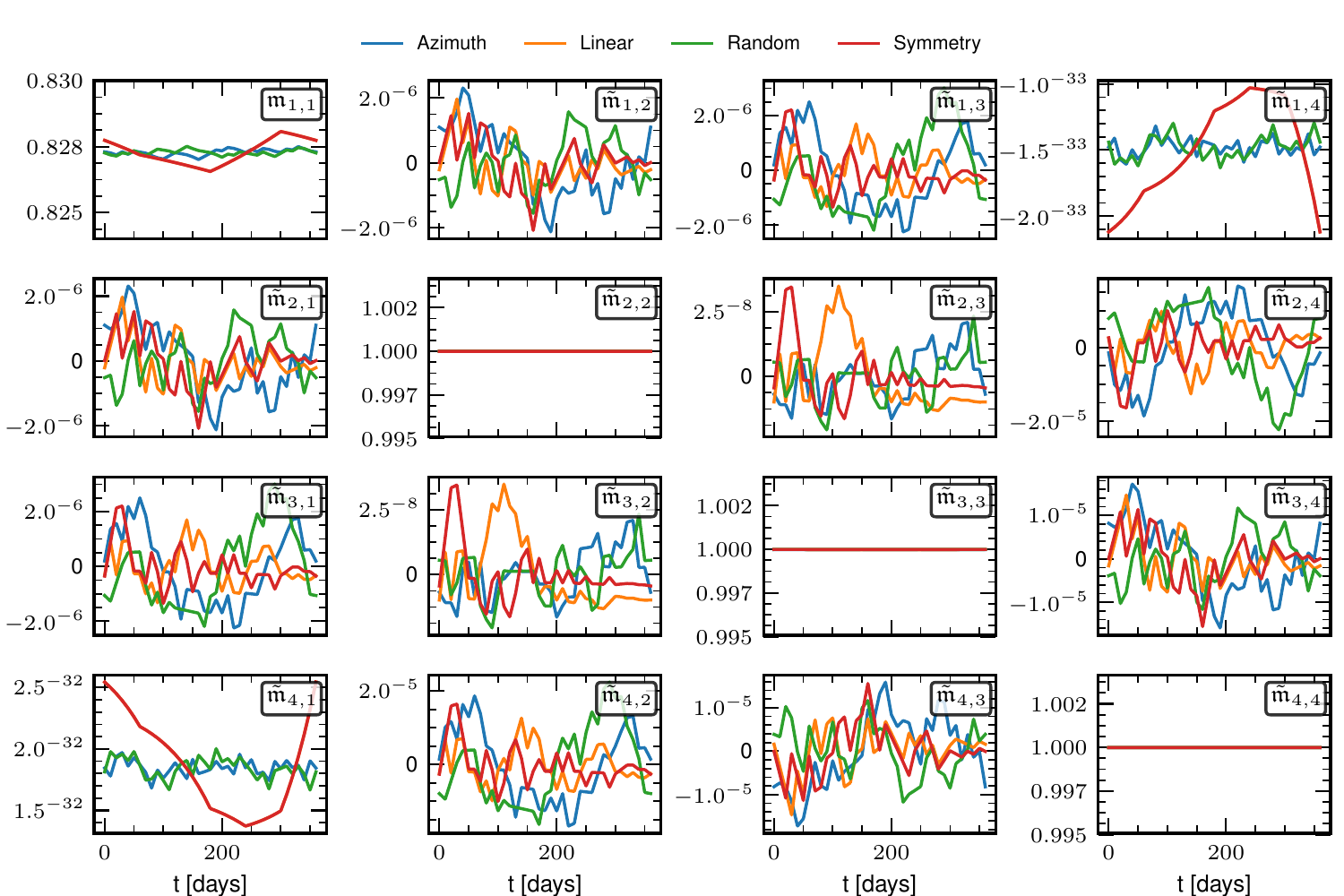}
    \caption{Time dependence of the Mueller matrix characterizing a GTC with conductor
    oxide for 5000 {\AA}. The time sequence covers a whole primary segment substitution
    sequence for the four segment substitution sequences (color coded). Note 
    that elements other than $\mathfrak{m}_{1,1}$ represent the reflection corrected
    value, i.e. $\tilde{\mathfrak{m}}_{i,j}=\mathfrak{m}_{i,j}/\mathfrak{m}_{1,1}$, for each time step.}
    \label{fig:oxidation_gtc_method}
\end{figure*}

\begin{equation*}
\mathfrak{M}_{\rm E-ELT}^{\rm ideal}=\left(
\begin{array}{cccc}
0.845048 & 1^{-18} & 1^{-19} & 1^{-35} \\
1^{-18} & 0.844903 & 1^{-17} & 1^{-18} \\
1^{-19} & 1^{-17} & 0.844903 & 1^{-17} \\
1^{-33} & 1^{-18} & 1^{-17} & 0.844758
\end{array}
\right).
\end{equation*}

In both cases, the ideal reflector has no instrumental polarization.

\subsection{The effect of oxidation}

The next ingredient we factor in is the formation of oxide on the conductor. 
To do so, we consider a time sequence for the GTC
telescope in which we replace a single segment of the primary
mirror at a time. By doing so, starting from a completely clean-ideal state, we
eventually reach a steady phase in which each individual segment is in a
different oxidation stage. We also discuss the different substitution
strategies described in Sect. 
\ref{subsect:replacement_sequences}
, with the aim
of emphasizing the importance of this substitution when doing
polarimetry in segmented primary mirrors. This is shown in Fig.
\ref{fig:oxidation_gtc_method} where this example follows the same inputs as
before, i.e. $\sim10^{7}$ rays at a wavelength of 500 nm. In addition, we change a
segment every 10 days so that in 360 days (approximately a year) all the segments
have been cleaned once.

In the non-ideal case all the cross-talk terms
except I$\leftrightarrow$V are non-zero. In particular, Q$\leftrightarrow$V
and U$\leftrightarrow$V terms reach $10^{-5}$, I$\leftrightarrow$Q and
I$\leftrightarrow$U terms are of the order of 
$10^{-6}$, and Q$\leftrightarrow$U term is of the order of $10^{-8}$.
This increase of the relative importance
of these terms is caused by the larger axial asymmetry induced by having a
primary mirror made of segments with different oxidation levels. 

\begin{figure*}
    \centering
    \includegraphics[width=0.85\textwidth]{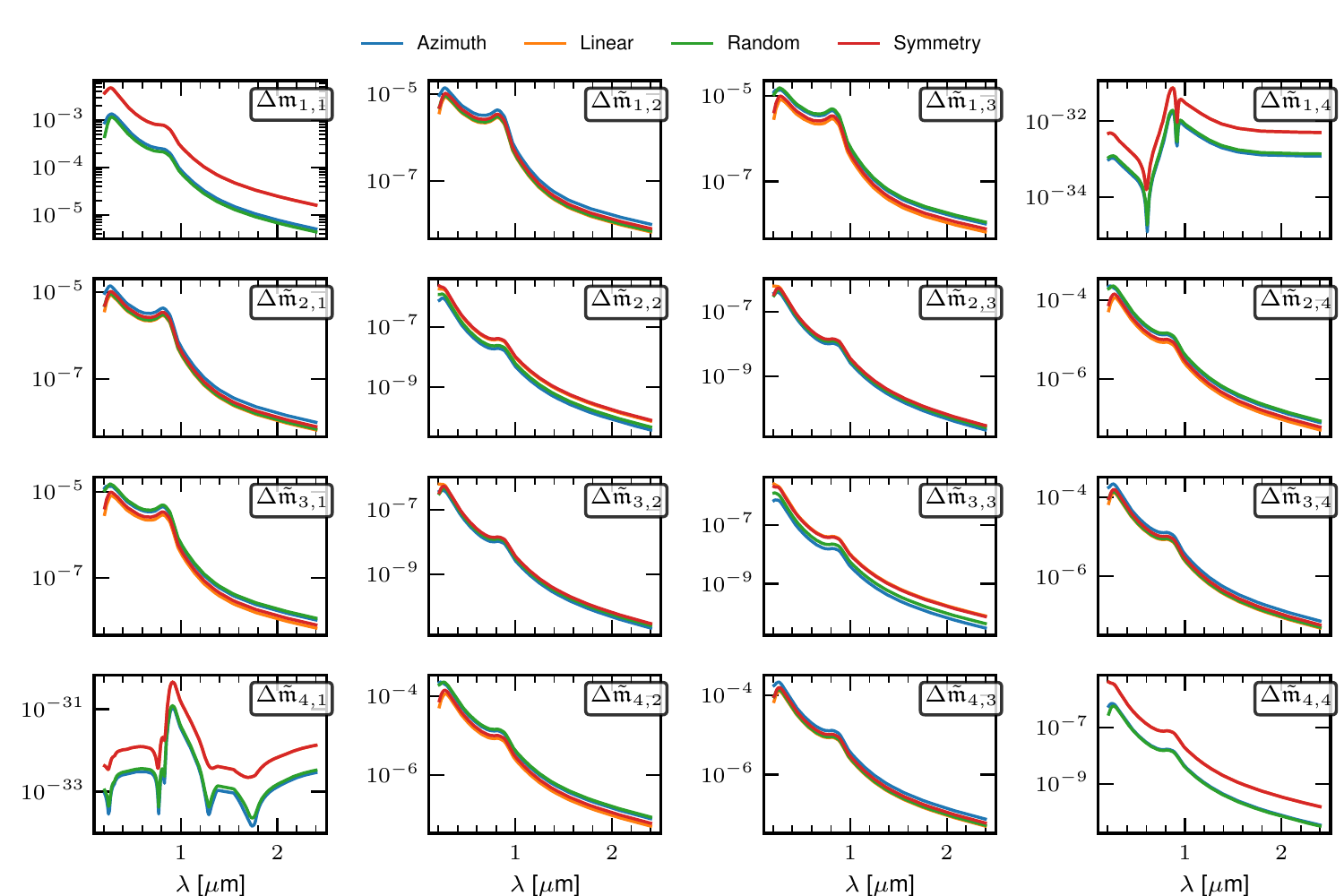}
    \caption{Maximum peak-to-peak variation of the Mueller matrix elements for a    complete GTC segment substitution sequence ($\sim$1 year) and its dependence with wavelength. Note that elements other than $\mathfrak{m}_{1,1}$ represent    the reflection corrected value, i.e. $\Delta\tilde{\mathfrak{m}}_{i,j}=\Delta\mathfrak{m}_{i,j}/\mathfrak{m}_{1,1}$, for each individual wavelength.}
    \label{fig:ptp_oxidation_gtc_method}
\end{figure*}

In principle, for typical scientific use cases, one can neglect 
the Q$\leftrightarrow$U term since this term
couples two Stokes parameters that are expected to be already small enough so that the
cross level stays well below the noise level. However, I$\leftrightarrow$Q and
I$\leftrightarrow$U terms must be taken carefully as linear polarization signals
of the order of $10^{-6}$ (and the specific value also depends on wavelength)
might be close to the pursued level of the polarimetric sensitivity
in the near future for large aperture telescopes. Similarly, strong linearly or 
circularly polarized sources can introduce
significant cross-talk by means of the Q$\leftrightarrow$V and
U$\leftrightarrow$V terms. There is no significant improvement on the total amount of
cross-talk induced for the different segment substitution sequences
considered here. Furthermore, it is not only the largest excursion values that matter
but also their variation with time, which can have an important impact on 
long term studies. In this sense, for all the substitution sequences
considered, they give rise to more or less similar cross-talk time
variation, i.e. for GTC case there seems to be no preferential way of
substituting primary mirror segments.

Concerning the transmittance of the system, we find two different behaviors. The
``linear'' and ``symmetry'' segment substitution sequences lead to a pronounced
time variation (of the order of $10^{-3}$) with a clear shape. This shape is
determined by the distance from the optical axis at which each exchanged segment
lies (see Fig. \ref{fig:muellersegmentdistance} for a reference on the
segment behavior as a function of the distance to the optical axis). 
The first six time steps correspond to the innermost segments, the next six
segments are the second closest segments to the optical axis and so on. 
Due to the larger incidence angle of segments far away from the optical axis, 
the effect over the transmittance of the system of exchanging an
innermost segment is smaller than exchanging an outermost one (see
Fig. \ref{fig:muellersegmentdistance}). It is significant
that this clear variation on the transmittance of the system has no counterpart
on the cross-talk terms. The ``azimuth'' and ``random'' substitution methods are
characterized by a flatter transmittance behavior all through the time sequence.

As we have previously seen, the polarimetric properties
depend upon the wavelength we are considering. In 
Fig. \ref{fig:ptp_oxidation_gtc_method} we explore the effect of the
oxidation dependency on wavelength 
by looking at the maximum peak-to-peak variation of each Mueller matrix element
 ($\Delta\mathfrak{m}_{i,j}$) on the time
sequence needed for a full mirror renovation with wavelength. For all
Mueller matrix elements the worst case scenario (the largest peak-to-peak variation)
is found for the blue part of the spectrum. It is significant that
I$\leftrightarrow$Q and I$\leftrightarrow$U cross-talk terms abruptly improve
approximately a $1/3$ fraction at around 0.4 $\mu$m and 
around an order of magnitude at around 1 $\mu$m. The other relevant
cross-talk terms (Q$\leftrightarrow$V, U$\leftrightarrow$V, and
Q$\leftrightarrow$U) present a more steady variation with wavelength with an
overall improvement when using the largest wavelengths
as compared to the blue part of the spectrum.

\begin{figure*}
    \centering
    \includegraphics[width=0.92\textwidth]{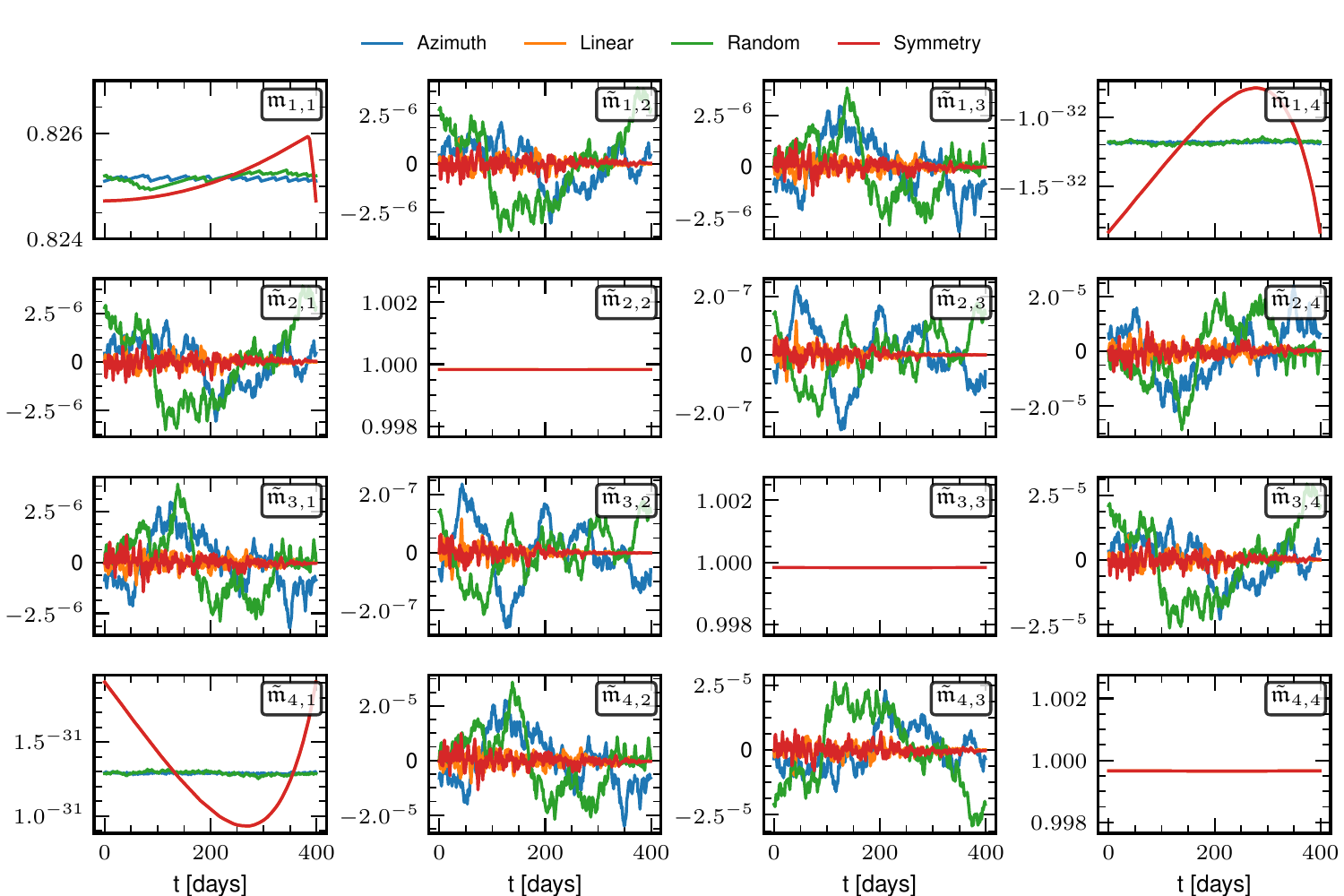}
    \caption{Same as in Fig. \ref{fig:oxidation_gtc_method} for \mbox{E-ELT}.}
    \label{fig:oxidation_eelt_method}
\end{figure*}

It is interesting to look to the \mbox{E-ELT} segment substitution sequences as it
gives rise to slightly different results, as displayed in Fig. \ref{fig:oxidation_eelt_method}.
The time sequence is obtained using $\sim 7\times10^{8}$
rays at 500 nm and changing two
segments everyday, so that the primary has been completely
renovated in 399 days. First, I$\leftrightarrow$Q and I$\leftrightarrow$U 
cross-talk terms reach $10^{-6}$,
Q$\leftrightarrow$U term reaches $10^{-7}$, while Q$\leftrightarrow$V and
U$\leftrightarrow$V can be of the order of $10^{-5}$. The peak amplitude of the 
Q$\leftrightarrow$U term is an order of magnitude larger than for the GTC case.
Second, and in clear contrast to GTC case, different substitution sequences lead
to huge differences in the actual value and the time evolution for the various
Mueller matrix elements. ``Linear'' and ``symmetry'' sequences lead to a smaller
peak-to-peak variation of the I$\leftrightarrow$Q, I$\leftrightarrow$U,
Q$\leftrightarrow$U, Q$\leftrightarrow$V, and U$\leftrightarrow$V as compared
with the ``azimuth'' and ``random'' substitution sequences. This
means that, in contrast to GTC, for \mbox{E-ELT} it might be interesting to follow
specific substitution sequences that minimize the variation with time of the
cross-talk terms of the optical system. Finally, diagonal elements do vary very
little presenting in the worst case (``linear'' and ``symmetry'') a 0.1\%
variation with time.

It is clear that the \mbox{E-ELT} case behaves differently to GTC. The reason is
that the number of segments is much larger and so the relative weight 
of each segment on the overall system is decreased. Consequently, there are more possibilities to
develop substitution sequences that minimize the time dependence of the instrumental
polarization by increasing the axial symmetry of the system. This is also
the reason why the variation with time of the transmittance of \mbox{E-ELT} is much
smaller than before, as each segment individually has a smaller effect on the
whole system and so the average state of all the segments is more homogeneous
than for the GTC. Eventually, there might be some segment substitution
sequences for \mbox{E-ELT} that are more suitable when using the telescope for polarimetric
studies, specially those that can minimize the I$\leftrightarrow$Q
and I$\leftrightarrow$U terms.

\begin{figure*}
    \centering
    \includegraphics[width=0.92\textwidth]{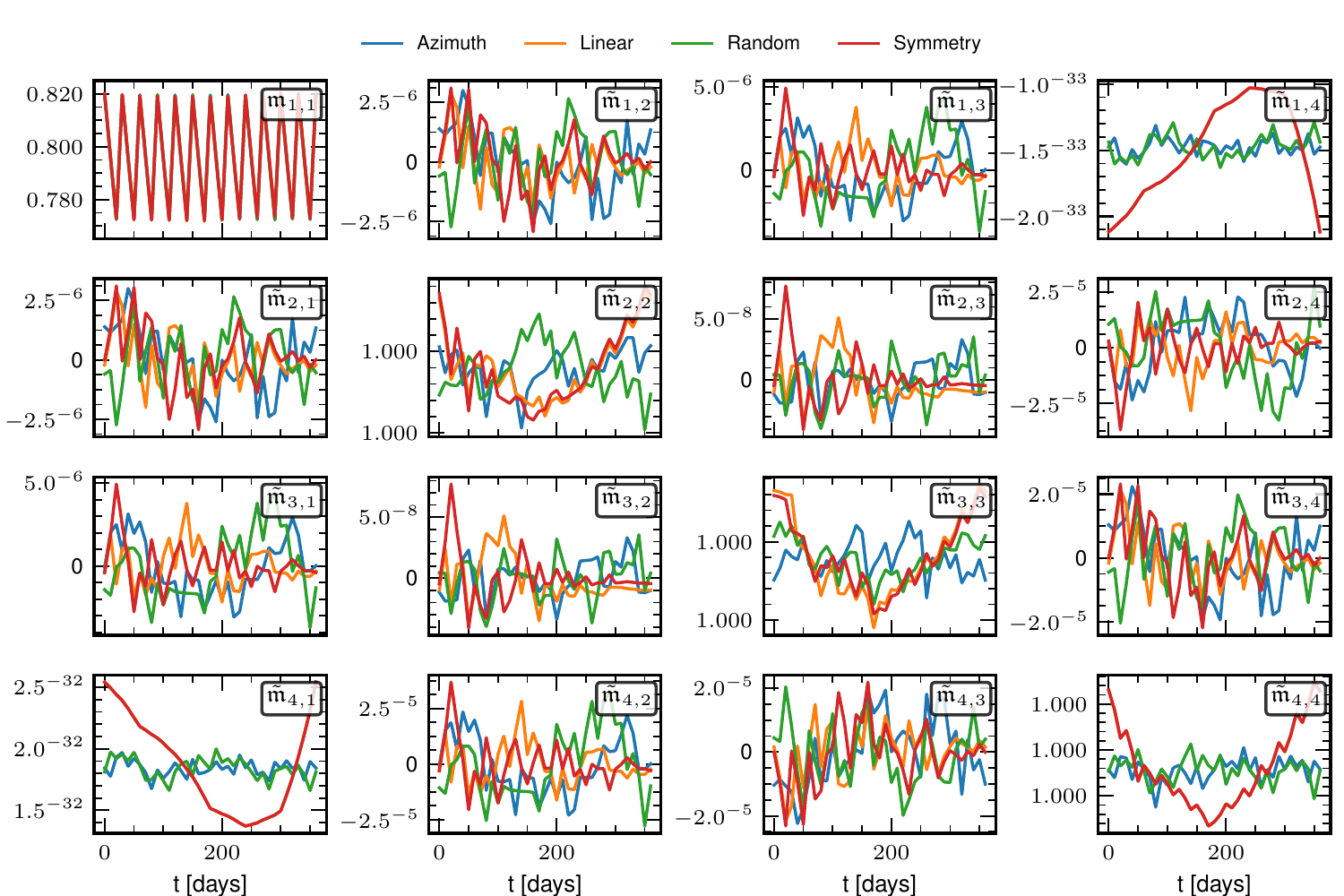}
    \caption{Same as in Fig. \ref{fig:oxidation_gtc_method} including oxidation and dust accumulation over the various segments.}
    \label{fig:oxidationdust_gtc_method}
\end{figure*}

\subsection{The effect of dust}

This section considers the complete model where we include both oxidation and dust 
accumulation on the segments of the primary mirror. The
model we employ for dust depends only on the rate of dust deposition ($\delta_{0}$) and
the efficiency ($\alpha$) and frequency in carbon dioxide snow cleaning. In
order to set $\delta_{0}$ and $\alpha$ to realistic values we have used the 
aging information of GTC segment mirrors as measured by the 
GTC team (private communication). We set fix $\delta_{0}=3\times10^{-3}$, $\alpha=90$\%,
and simulate that a snow cleaning procedure takes place every 30 days. 
Here, we limit the example to the GTC case because we find no significant 
effect on the global Mueller matrix when dust is included. Again, we have used
500 nm wavelength and $\sim10^{7}$ rays and the results are displayed
in Fig. \ref{fig:oxidationdust_gtc_method}.
The main difference with the case with oxidation only (see Fig.
\ref{fig:oxidation_gtc_method}) is found in 
the transmittance of the optical system (see element $\mathfrak{m}_{1,1}$ in Fig. 
\ref{fig:oxidationdust_gtc_method}), whose shape is dominated by the attenuation
as dust is accumulated (the probability of a ray hitting dust and thus being 
discarded increases) and a sudden improvement of the whole system takes place as
a the mirror is cleaned (i.e. it mainly affects to the whole system at once). Actually, cross-talk terms
(I$\leftrightarrow$Q, I$\leftrightarrow$U, Q$\leftrightarrow$U,
Q$\leftrightarrow$V, and U$\leftrightarrow$V) show no significant variation
(neither in values nor in shape) as compared with the case in absence of dust (Fig.
\ref{fig:oxidation_gtc_method}). The reason for this behavior is that
the dust model has a significant degree of axial symmetry. Even
though dust has a relative large effect on each segment (as it is clear from the
saw-shape transmittance profiles), its effect is quite homogeneous in azimuth
and so it has very limited effect on the cross-talk terms. It is to be noticed
that this result might be very model dependent. The dust model here employed
is quite simple as compared to the dust distribution one might encounter in
reality. For instance, one might expect (as it is actually observed at GTC) that
the inner mirrors get more dust than the outer ones. It is so because gravity
helps maintaining the external mirrors cleaner than the internal ones. Also, the
presence of wind micro-currents leads to some axial symmetry loss in the
accumulation of dust. Nevertheless, since the latter is a minor effect, one
expects a significantly symmetric dust deposition over the primary mirror and in
that case, according to the results here found, one would not expect a dramatic
change in the cross-talk terms behavior. 

\section{Conclusion}

In summary we have presented a numerical code 
\footnote{\url{https://github.com/apy-github/segmented_mirrors}}
that estimates the induced instrumental 
polarization of an optical system. It is done by means of ray tracing and using 
the Fresnel coefficients
and the Stokes formalism. Our focus is to quantify
the effect that the axial symmetry breaking
around the optical axis has on the polarimetric behavior of the optical system
with a segmented primary mirror and an ideal secondary mirror. 
The usage of segmented primary mirrors allows the sequential
replacement of segments whose main effect over the polarimetric behavior of the
optical system is an important decrease of the axial symmetry. Here, we have
considered the appearance of oxide over the mirror conductor as well as the
accumulation of dust. We have found that the different time evolution of these
two aging sources (mostly oxidation) leads to the appearance of cross-talk terms
that can reach the order of magnitude of the expected polarization signals
($10^{-5}$). The fact that dust has little impact on the polarimetric behavior
of the system can be a consequence of the simple model we have used in this work. 
A detailed-accurate model of dust is extremely difficult to achieve and it is out
of the scope for the present work, but even in its current model we think the
results are significant for two main reasons: 1- even simple, the model for dust
here considered can produce differential polarimetric behavior among the various segments.
2- Employing more complex dust models will not change the symmetry of the problem and
thus the overall impact on the polarimetric behavior of the system will be largely reduced
. The induced instrumental polarization is wavelength dependent and it is
more relevant in the blue part of the spectrum than in the red, where their effect is largely
minimized (this effect depends on the refraction index of the
conductor being used). An important result is that one can potentially find
substitution sequences in telescopes with primary mirrors with a sufficiently
small segment-to-mirror area fraction (in this case \mbox{E-ELT}) that clearly reduce the induced 
polarization terms. We leave the study of these substitution sequences 
for a future work.

\begin{backmatter}
\bmsection{Funding}
This work has received funding from the European Research Council (ERC) under the
European Union's Horizon 2020 research and innovation programme (SUNMAG, grant
agreement 759548), the State Research Agency (AEI) of the Spanish Ministry of
Science, Innovation and Universities (MCIU) and the European Regional Development
Fund (FEDER) (PGC2018-097611-A-I00 and PGC2018-102108-B-I00).

\bmsection{Acknowledgments}
We express our appreciation to Antonio Luis Cabrera Lavers and Manuela Abril
Abril for their assistance with GTC procedures and measurements. This research
has made use of NASA's Astrophysics Data System Bibliographic Services. We
acknowledge the community effort devoted to the development of the following
open-source packages that were used in this work: \texttt{numpy}
(\texttt{numpy.org}) \cite{numpy2020}, \texttt{matplotlib}
(\texttt{matplotlib.org})\cite{hunter2007}, and \texttt{scipy}
(\texttt{scipy.org})\cite{scipy2020}.

\bmsection{Disclosures} The authors declare no conflicts of interest.

\bmsection{Data availability} The data that support the findings of this study are available from the corresponding author upon reasonable request.
\end{backmatter}

\bibliography{sample}

\begin{thebibliography}{10}
\newcommand{\enquote}[1]{``#1''}

\bibitem{gilmozzi2007}
R.~{Gilmozzi} and J.~{Spyromilio}, \enquote{{The European Extremely Large
  Telescope (E-ELT)},} {\protect\JournalTitle{The Messenger}} \textbf{127}, 11
  (2007).

\bibitem{sanders2013}
G.~H. {Sanders}, \enquote{{The Thirty Meter Telescope (TMT): An International
  Observatory},} {\protect\JournalTitle{Journal of Astrophysics and Astronomy}}
  \textbf{34}, 81--86 (2013).

\bibitem{brun2017}
A.~S. {Brun} and M.~K. {Browning}, \enquote{{Magnetism, dynamo action and the
  solar-stellar connection},} {\protect\JournalTitle{Living Reviews in Solar
  Physics}} \textbf{14}, 4 (2017).

\bibitem{martinezgonzalez2015}
M.~J. {Mart{\'\i}nez Gonz{\'a}lez}, A.~{Asensio Ramos}, R.~{Manso Sainz},
  R.~L.~M. {Corradi}, and F.~{Leone}, \enquote{{Constraining the shaping
  mechanism of the Red Rectangle through the spectro-polarimetry of its central
  star},} {\protect\JournalTitle{\aap}} \textbf{574}, A16 (2015).

\bibitem{kochukhof2016}
O.~{Kochukhov}, \emph{{Doppler and Zeeman Doppler Imaging of Stars}} (2016),
  vol. 914, p. 177.

\bibitem{carnerero2017}
M.~I. {Carnerero}, C.~M. {Raiteri}, M.~{Villata}, J.~A. {Acosta-Pulido}, V.~M.
  {Larionov}, P.~S. {Smith}, F.~{D'Ammando}, I.~{Agudo}, M.~J. {Ar{\'e}valo},
  R.~{Bachev}, J.~{Barnes}, S.~{Boeva}, V.~{Bozhilov}, D.~{Carosati},
  C.~{Casadio}, W.~P. {Chen}, G.~{Damljanovic}, E.~{Eswaraiah}, E.~{Forn{\'e}},
  G.~{Gantchev}, J.~L. {G{\'o}mez}, P.~A. {Gonz{\'a}lez-Morales}, A.~B.
  {Gri{\~n}{\'o}n-Mar{\'\i}n}, T.~S. {Grishina}, M.~{Holden}, S.~{Ibryamov},
  M.~D. {Joner}, B.~{Jordan}, S.~G. {Jorstad}, M.~{Joshi}, E.~N. {Kopatskaya},
  E.~{Koptelova}, O.~M. {Kurtanidze}, S.~O. {Kurtanidze}, E.~G. {Larionova},
  L.~V. {Larionova}, G.~{Latev}, C.~{L{\'a}zaro}, R.~{Ligustri}, H.~C. {Lin},
  A.~P. {Marscher}, C.~{Mart{\'\i}nez-Lombilla}, B.~{McBreen}, B.~{Mihov},
  S.~N. {Molina}, J.~W. {Moody}, D.~A. {Morozova}, M.~G. {Nikolashvili},
  K.~{Nilsson}, E.~{Ovcharov}, C.~{Pace}, N.~{Panwar}, A.~{Pastor Yabar}, R.~L.
  {Pearson}, F.~{Pinna}, C.~{Protasio}, N.~{Rizzi}, F.~J. {Redondo-Lorenzo},
  G.~{Rodr{\'\i}guez-Coira}, J.~A. {Ros}, A.~C. {Sadun}, S.~S. {Savchenko},
  E.~{Semkov}, L.~{Slavcheva-Mihova}, N.~{Smith}, A.~{Strigachev}, Y.~V.
  {Troitskaya}, I.~S. {Troitsky}, A.~A. {Vasilyev}, and O.~{Vince},
  \enquote{{Dissecting the long-term emission behaviour of the BL Lac object
  Mrk 421},} {\protect\JournalTitle{\mnras}} \textbf{472}, 3789--3804 (2017).

\bibitem{berdyugina2008}
S.~V. {Berdyugina}, A.~V. {Berdyugin}, D.~M. {Fluri}, and V.~{Piirola},
  \enquote{{First Detection of Polarized Scattered Light from an Exoplanetary
  Atmosphere},} {\protect\JournalTitle{\apjl}} \textbf{673}, L83 (2008).

\bibitem{kiang2018}
N.~Y. {Kiang}, S.~{Domagal-Goldman}, M.~N. {Parenteau}, D.~C. {Catling},
  Y.~{Fujii}, V.~S. {Meadows}, E.~W. {Schwieterman}, and S.~I. {Walker},
  \enquote{{Exoplanet Biosignatures: At the Dawn of a New Era of Planetary
  Observations},} {\protect\JournalTitle{Astrobiology}} \textbf{18}, 619--629
  (2018).

\bibitem{capitani1989}
C.~{Capitani}, E.~{Landi Degl'Innocenti}, F.~{Cavallini}, G.~{Ceppatelli},
  M.~{Landi Degl'Innocenti}, M.~{Landolfi}, and A.~{Righini},
  \enquote{{Polarization properties of a `Zeiss-type' coelostat: The case of
  the solar tower in Arcetri},} {\protect\JournalTitle{\solphys}} \textbf{120},
  173--191 (1989).

\bibitem{collados2010}
M.~{Collados}, F.~{Bettonvil}, L.~{Cavaller}, I.~{Ermolli}, B.~{Gelly},
  C.~{Grivel-Gelly}, A.~{P{\'e}rez}, H.~{Socas-Navarro}, D.~{Soltau}, and
  R.~{Volkmer}, \enquote{{European Solar Telescope: project status},} in
  \emph{Ground-based and Airborne Telescopes III,}  vol. 7733 of
  \emph{\procspie} (2010), p. 77330H.

\bibitem{telesco2003}
C.~M. {Telesco}, D.~{Ciardi}, J.~{French}, C.~{Ftaclas}, K.~T. {Hanna}, D.~B.
  {Hon}, J.~H. {Hough}, J.~{Julian}, R.~{Julian}, M.~{Kidger}, C.~C. {Packham},
  R.~K. {Pina}, F.~{Varosi}, and R.~G. {Sellar}, \enquote{{CanariCam: a
  multimode mid-infrared camera for the Gran Telescopio CANARIAS},} in
  \emph{Instrument Design and Performance for Optical/Infrared Ground-based
  Telescopes,}  vol. 4841 of \emph{Society of Photo-Optical Instrumentation
  Engineers (SPIE) Conference Series} M.~{Iye} and A.~F.~M. {Moorwood}, eds.
  (2003), pp. 913--922.

\bibitem{eikenberry2016}
S.~S. {Eikenberry}, S.~N. {Raines}, R.~D. {Stelter}, A.~{Garner},
  Y.~{Dallilar}, K.~{Ackley}, J.~G. {Bennett}, C.~H. {Murphey}, P.~{Miller},
  D.~{Tooke}, L.~{Williams}, B.~{Chinn}, S.~A. {Mullin}, S.~L. {Schofield},
  C.~D. {Warner}, F.~{Varosi}, B.~{Zhao}, S.~A. {Eikenberry}, C.~{Vega}, H.~V.
  {Donoso}, J.~{Sabater}, J.~M. {G{\'o}mez}, J.~{Torra}, J.~{Rosich Minguell},
  F.~{Garz{\'o}n L{\'o}pez}, N.~{Cardiel}, J.~{Gallego Maestro},
  A.~{Mar{\'\i}n-Franch}, J.~{Galipienzo}, M.~{\'A}. {Carrera Astigarraga},
  G.~J. {Fitzgerald}, I.~{Prees}, T.~M. {Stolberg}, P.~A. {Kornik}, A.~N.
  {Ramaprakash}, M.~P. {Burse}, S.~P. {Punnadi}, and P.~{Hammersley},
  \enquote{{MIRADAS for the Gran Telescopio Canarias},} in \emph{Ground-based
  and Airborne Instrumentation for Astronomy VI,}  vol. 9908 of \emph{Society
  of Photo-Optical Instrumentation Engineers (SPIE) Conference Series} C.~J.
  {Evans}, L.~{Simard}, and H.~{Takami}, eds. (2016), p. 99081L.

\bibitem{jennings2002}
J.~K. {Jennings} and B.~T. {Landesman}, \emph{{Polarization effects from large
  non-optically flat segmented mirrors}} (2002), vol. 4489 of \emph{Society of
  Photo-Optical Instrumentation Engineers (SPIE) Conference Series}, pp.
  89--99.

\bibitem{anche2018}
R.~M. {Anche}, A.~K. {Sen}, G.~C. {Anupama}, K.~{Sankarasubramanian}, and
  W.~{Skidmore}, \enquote{{Analysis of polarization introduced due to the
  telescope optics of the Thirty Meter Telescope},}
  {\protect\JournalTitle{Journal of Astronomical Telescopes, Instruments, and
  Systems}} \textbf{4}, 018003 (2018).

\bibitem{anche2019}
R.~M. {Anche}, G.~C. {Anupama}, S.~{Sriram}, and K.~{Sankarasubramanian},
  \enquote{{Polarization effects due to the segmented primary mirror of the
  Thirty Meter Telescope},} in \emph{American Astronomical Society Meeting
  Abstracts \#233,}  vol. 233 of \emph{American Astronomical Society Meeting
  Abstracts} (2019), p. 437.01.

\bibitem{yun2011}
G.~{Yun}, K.~{Crabtree}, and R.~A. {Chipman}, \enquote{{Three-dimensional
  polarization ray-tracing calculus I: definition and diattenuation},}
  {\protect\JournalTitle{\ao}} \textbf{50}, 2855 (2011).

\bibitem{born1999}
M.~{Born} and E.~{Wolf}, \emph{{Principles of Optics}} (1999).

\bibitem{numpy2020}
C.~R. Harris, K.~J. Millman, S.~J. van~der Walt, R.~Gommers, P.~Virtanen,
  D.~Cournapeau, E.~Wieser, J.~Taylor, S.~Berg, N.~J. Smith, R.~Kern, M.~Picus,
  S.~Hoyer, M.~H. van Kerkwijk, M.~Brett, A.~Haldane, J.~Fernández~del Río,
  M.~Wiebe, P.~Peterson, P.~Gérard-Marchant, K.~Sheppard, T.~Reddy,
  W.~Weckesser, H.~Abbasi, C.~Gohlke, and T.~E. Oliphant, \enquote{Array
  programming with {NumPy},} {\protect\JournalTitle{Nature}} \textbf{585},
  357–362 (2020).

\bibitem{hunter2007}
J.~D. Hunter, \enquote{Matplotlib: A 2d graphics environment,}
  {\protect\JournalTitle{Computing in Science \& Engineering}} \textbf{9},
  90--95 (2007).

\bibitem{scipy2020}
P.~Virtanen, R.~Gommers, T.~E. Oliphant, M.~Haberland, T.~Reddy, D.~Cournapeau,
  E.~Burovski, P.~Peterson, W.~Weckesser, J.~Bright, S.~J. {van der Walt},
  M.~Brett, J.~Wilson, K.~J. Millman, N.~Mayorov, A.~R.~J. Nelson, E.~Jones,
  R.~Kern, E.~Larson, C.~J. Carey, {\.I}.~Polat, Y.~Feng, E.~W. Moore,
  J.~{VanderPlas}, D.~Laxalde, J.~Perktold, R.~Cimrman, I.~Henriksen, E.~A.
  Quintero, C.~R. Harris, A.~M. Archibald, A.~H. Ribeiro, F.~Pedregosa, P.~{van
  Mulbregt}, and {SciPy 1.0 Contributors}, \enquote{{{SciPy} 1.0: Fundamental
  Algorithms for Scientific Computing in Python},}
  {\protect\JournalTitle{Nature Methods}} \textbf{17}, 261--272 (2020).

\end{thebibliography}

\bibliographyfullrefs{sample}

\end{document}